\documentclass{emulateapj}
\newcommand{\Msun}{\mbox{M$_{\odot}$}}
\newcommand{\Rsun}{\mbox{R$_{\odot}$}}
\newcommand{\Mjup}{\mbox{M$_{\rm Jup}$}}
\newcommand{\Rjup}{\mbox{R$_{\rm Jup}$}}

\newcommand{\ms}  {\mbox{m \ s$^{-1}$}}

\slugcomment{To appear in the Astrophysical Journal Letters}

\shorttitle{Ellipsoidal Variations in HAT-P-7}
\shortauthors{Welsh et al.}

\begin{document}

\title{ The Discovery of Ellipsoidal Variations in the
{\it{Kepler}} Light Curve of HAT-P-7}

\author{William F. Welsh, Jerome A. Orosz}
\affil{Astronomy Department, San Diego State University,
San Diego, CA 92182 USA}
\email{wfw@sciences.sdsu.edu}

\author{Sara Seager}
\affil{Massachusetts Institute of Technology, MA 02139 USA}

\author{Jonathan J. Fortney }
\affil{University of California, Santa Cruz, CA 95064 USA}

\author{Jon Jenkins}
\affil{SETI Institute, Mountain View, CA 94043, USA}

%%%% \and

\author{Jason F. Rowe\footnote{NASA Postdoctoral Program Fellow}, 
David Koch, and William J. Borucki}
\affil{NASA Ames Research Center, Moffett Field, CA 94035 USA}

%.............................................................
\begin{abstract}

We present an analysis of the early {\it{Kepler}} observations
of the previously discovered transiting planet HAT-P-7b. 
The light curve shows the transit of the star, the occultation of
the planet, and the orbit phase-dependent light from the planet.
In addition, phase-dependent light from the star is present, known
as ``ellipsoidal variations''. The very nearby planet (only 4 stellar 
radii away) gravitationally distorts the star and results in
a flux modulation twice per orbit. The ellipsoidal variations can 
confuse interpretation of the planetary phase curve if not 
self-consistently included in the modeling. We fit the light 
curve using the Roche potential approximation and derive improved 
planet and orbit parameters. 
\end{abstract}

\keywords{ binaries: eclipsing}
%.............................................................

\section{Introduction}

{\it{Kepler}} is a reconnaissance mission to obtain time-series 
optical photometry of $\sim$150,000 stars in order to determine 
characteristics of Earth-size and larger extrasolar planets: 
frequencies, sizes, orbital distributions, and correlations with 
the properties of the host stars \citep{Borucki10}. To achieve 
these goals, {\it{Kepler}} requires exceptional long-term 
photometric stability and precision \citep{Koch10}. In addition 
to the discovery aspect of the mission, this unprecedented 
photometric capability provides exquisite observations of a host 
of astrophysical objects, including the previously known 
extrasolar planets in the {\it{Kepler}} field of view. In this 
{\it Letter}, we examine the early {\it{Kepler}} observations 
of the planet HAT-P-7b.

Discovered via the {\it HATNet} project, the transiting planet 
HAT-P-7b revolves in a tight circular orbit (a=0.038 AU, 
P=2.20473~d) around a bright (V=10.5 mag) F6 star \citep{Pal08}. 
The proximity to its 6350~K host star \citep{Pal08} means the 
planet is highly irradiated, resulting in very high temperatures 
($\sim$2140~K), making it an extreme pM-type planet 
\citep{Fortney08}. Using observations of the Rossiter-McLaughlin 
effect, 
\citet{Narita09} and \citet{Winn09} find that the planet's 
orbital axis is extremely tilted compared to the star's spin 
axis, and probably even retrograde, 
implying an interesting formation and orbital 
evolution history. Additionally, the radial velocities exhibit
an acceleration, suggesting the presence of another body in the 
system, perhaps responsible for the tilted orbit \citep{Winn09}.

Observations of HAT-P-7 during the 10 days of commissioning of 
the {\it{Kepler}} photometer revealed the presence of an occultation 
(also known as a ``secondary eclipse'') as the planet passes 
behind the star \citep{Borucki09}. These data also show the phase 
``reflected'' light from the planet, a result of both scattered 
and thermal emission. Additional observations of HAT-P-7 have 
been obtained, and we present these in \S2. In \S3 we describe 
our modeling method, with particular emphasis on the ellipsoidal 
variations from the star. In \S4 we present and discuss our 
findings.

% ................................
\section{Observations}
HAT-P-7 was monitored continuously for 33.5~d during the
2009 May 13--Jun 15 ``Quarter 1'' (Q1) epoch in short-cadence 
mode. The 42 {\it{Kepler}} CCDs were read out every 6~s and 
co-added on-board to achieve approximately 1~min sampling 
cadence. The photometer has no shutter, so an overscan region 
is used to remove the effects of smearing during readout.
In 6~s exposures HAT-P-7 saturates the CCD; however, because 
the {\it{Kepler}} photometer is such a stable platform, this 
does not hamper the relative precision and superb photometry 
is possible. For details on the design and performance of 
the {\it{Kepler}} photometer, see \citet{Koch10} and 
\citet{Jenkins10}.

Fifteen complete transits of HAT-P-7 were observed, and after 
removing 21 cosmic rays via a +4 sigma rejection from a 30-min 
wide running median, 49,023 measurements remained. The 
background-subtracted light curve exhibited a gradual drop in 
counts (0.13\%), suspected to be due to a drift in focus coupled 
with a fixed extraction aperture. This trend was removed using a 
cubic polynomial (masking out the transits). Because we are 
interested in phenomena on the orbital timescale (2.2 d), we 
chose this modest detrending to leave in as much power as 
possible on orbital timescales. 
Fluxes were normalized at phase 0.5 (mid occultation) 
because the planet is hidden at this phase and we see starlight 
alone. Short-timescale systematic calibration features are 
present \citep{Gilliland10}, but they does not affect the 
analysis other than being a noise term. Uncertainties were 
estimated by taking the median of the rms deviations in 30-min 
bins, resulting in 150 ppm per 1-min datum.

The detrended and phase-folded light curve is shown in Fig.~1, 
where the upper panel includes our fit to the transit, and the 
lower panel highlights the eclipse of the planet and the 
phase-dependent light from the planet and star. Notice that the 
maximum does not occur just outside occultation as one would 
expect for simple ``reflection'' from the planet. Two maxima 
occur 0.15-0.20 away in phase, a result of the ``ellipsoidal 
variation'' of the star's light, as discussed in \S4. 

The absolute timing is still preliminary in this early version 
of the data calibration pipeline, but relative timing precision 
is reliable (though without the modest BJD correction, which is 
unimportant for our investigation over the 33.5~d time series). 
So we do not report the value for the epoch of transit T0, 
though of course it is a parameter in the model fitting. 
Also, the period measured is based only on these 33.5~d, 
so the precision is not as high as would be if other epochs 
many cycles away were included.

% .................................................
\section{Modeling}
We employ the ELC code of \citet{Orosz00} to model the transit, 
occultation, and phase-varying light from the planet and star. 
The code simultaneously fits the photometry along with several 
observational parameters: the radial velocity K, and the mass 
and radius of the star. The K amplitude was taken from 
\citet{Winn09}, and the mass and radius from the asteroseismology 
analysis of the {\it{Kepler}} data \citep{Christensen10}.
These parameters are not fixed; rather the models are started 
and steered toward them via a chi-square penalty for deviations. 
Markov chain Monte Carlo and genetic algorithms were used to 
search parameter space, find the global chi-square minimum, and 
determine confidence intervals of the fitted parameters.

The analytic model of \citet{Gimenez06} for a spherical star 
and planet is not sufficient to model the phase variations. 
Therefore ELC is used in its full numerical mode: 
the star and planet are tiled in a fine grid, and the intensity 
and velocity from each tile is summed to give the light curve 
and radial velocities. Limb and gravity darkening are included, 
and the gravitational distortions are modeled assuming a 
standard Roche potential.
% As discussed in the next section, the non-sphericity of the star 
% manifests itself by ellipsoidal variations in the light curve.
We employed a blackbody approximation evaluated at a wavelength
of 6000~\AA, and a hybrid 
method\footnote{
ELC can also fully employ stellar atmosphere intensities, where 
no parameterized limb darkening is used. But preliminary tests 
gave significantly worse fits unless the stellar temperature was 
allowed to be several thousand degrees hotter. The cause seems 
to be related to the very wide {\it{Kepler}} bandpass and the 
lack of freedom to adjust the limb darkening.} 
where model atmospheres are used to determine the intensities at 
the normal for each tile and a parameterized limb darkening law 
is used for other angles; the model is then filtered through 
the {\it Kepler} spectral response function
(spanning roughly 4250--8950~\AA, peaking at 5890~\AA \ with 
a mean wavelength of 6400~\AA \ --- see \citet{Koch10} and 
{\citet{VanCleve09}.
The two methods give essentially the same results, with the 
exception of a higher albedo from the hybrid models.
Interestingly, the blackbody models yielded significantly 
lower chi-square values, so we quote the blackbody model values 
in this work. This needs to be kept in mind when interpreting 
the temperatures and albedos estimates given in \S4.
We use a 2-parameter logarithmic limb darkening law and adopt an 
eccentricity of zero consistent with the radial velocities and phase 
of occultation. We assume the planet is tidally locked in synchronous 
rotation. For more details on using ELC to model exoplanet data see 
\citet{Wittenmyer05}.

Following the prescription of \citet{Wilson90}, the light from 
the planet is modeled as the sum of an 
isothermal component (with temperature $T_{p}$ that essentially 
adds a constant flux at all phases outside of occultation), and a 
``reflection'' component on the day hemisphere. 
The local temperature is given by
$T^{4} = T^{4}_{p} \times 
\left[ 1 + A_{bol} \frac{F_{*}}{F_{p}} \right]$
which comes from assuming the fluxes and temperature are coupled 
by the Stefan-Boltzmann law. The bolometric albedo $A_{bol}$, 
also known as the ``heat albedo'' is the ratio of re-radiated
to incident energy, and should not be confused with the Bond albedo.
For stars, a radiative atmosphere has $A_{bol}$=1 (local energy 
conservation) while for convective atmospheres a value of 0.5 is 
appropriate (half the energy is radiated, half gets redistributed)
--- see \citet{Kallrath99} for a full description of the method. 
In our model we allow $A_{bol}$ and the temperature ratio 
$T_{p}/T_{*}$ to be free parameters, keeping $T_{*}$ fixed at 
6350~K \citep{Pal08}.
The term in brackets is the local reflection factor $\mathcal{R}$, 
equal to the ratio of the total radiated flux (internal + 
re-radiated) to internal flux. Thus $\mathcal{R} \ge 1$, with 
$\mathcal{R}=1$ on the night side. For an isolated planet, $T_{p}$ 
is very low (e.g. \citet{Burrows06} assume 50~K), but for an 
irradiated planet, much of the incident energy is eventually 
re-distributed, bringing the night-time temperature up to 
much higher temperatures.

In the ELC model, the local emitted flux is completely 
dependent on the local temperature.
The local temperature depends on the mean effective temperature,
the local gravity, and irradiation. The irradiation term 
is the sum over all the visible tiles on the star as seen from
each tile on the planet, and includes the ellipsoidal shape of 
the star, gravity darkening, limb darkening, and penumbra 
correction (accounting for the fact that parts of the star are 
not visible because they are blocked by the local horizon). 
Note that given the close proximity of the planet to the star, 
no part of the planet sees a full hemisphere of the star.
There is no explicit scattering term in the reflection: 
the incident radiation heats up the planet and is always 
re-emitted locally. Scattered light is implicitly accounted 
for by allowing the global temperature $T_{p}$ to be de-coupled 
from the day-side temperature.

The ELC model provides a good fit to the observations. 
The chi square is 57,225 for 49,016 degrees of freedom 
(reduced $\chi^{2}_{\nu}$=1.167).
Uncertainties on the parameters were boosted by a factor 
of $\sqrt{\chi^{2}_{\nu}}$ (=8\%) to account for the formally 
high $\chi^{2}_{\nu}$, which we attribute to systematic 
non-Gaussian noise. The values of the parameters are listed in 
Table~1.

% ................................
\section{Results and Discussion}

The 1.8 $\Mjup$ planet orbiting only 4.1 stellar radii 
from its host star induces a tidal distortion on the star, 
changing its shape from oblate (not spherical, due to its 
rotation) to a more triaxial shape, with the longest axis along 
the direction toward the planet and the shortest axis 
perpendicular to the orbital plane.\footnote{ More exactly, in a 
Roche potential there are 4 distinct radii: towards the companion, 
perpendicular to the orbit plane, along the direction of motion, 
and away from the companion.} This shape causes the well-known 
``ellipsoidal variation'' effect seen in binary stars: 
a modulation in light at half the orbital period, with maxima 
at phases near 0.25 and 0.75, and unequal minima at phases 0.0 
and 0.5.
The ellipsoidal variation is primarily a geometrical 
(projected surface area) effect whose relative amplitude depends 
on the mass ratio and the inclination of the binary system. 
Given the tight constraint on the inclination from the eclipses, 
they may provide some limits on the mass ratio, and hence the 
mass of the planet. In the optical, where the phase-dependent 
planet-to-star flux ratio is small, the presence of the stellar 
ellipsoidal variation becomes significant. Neglecting its 
contribution can lead to a confused interpretation of the phase 
curve and thus an incorrect measurement of the albedo and 
phase-dependent scattered/thermal emission.

The presence of ellipsoidal variations in exoplanetary systems 
was anticipated by \citet{Loeb03} and \citet{Drake03}, and
\citet{Pfahl08} present a detailed theoretical investigation of 
the tidal force on the star by the planet.
Figs.~1 and 3 show the first detection of the effect in an 
exoplanet system.
In Fig.~3 we show the light curve binned to 30~min, along with 
the best-fit model decomposition. The dotted curve shows the 
stellar-only ellipsoidal light curve while the dashed curve is 
the planet-only light curve (offset vertically). The solid curve 
is the sum of the two and equals our best-fit model. The 
planet-only model can match the light curve only very near phase 
0.5; it is a very poor fit to the data at other phases, 
indicating the need for the ellipsoidal variation component. 
The amplitude of the ellipsoidal component is 37.3~ppm, 
detectable only because of {\it{Kepler's}} high-precision 
photometric capability.

Ellipsoidal variations arise as a consequence of gravity on a 
luminous fluid body. Within the Roche framework, its amplitude is 
exactly known if the mass ratio, inclination, and stellar radius 
are known. However, since in exoplanet systems the star is not 
expected to be in synchronous rotation, the Roche potential is an 
approximation, albeit a good one since the maximum tidal 
distortion $\Delta R/R$ is only $10^{-4}$. However, we stress 
that the model presented here is only a starting point, based on 
the well-developed and successful Roche model. Effects such as 
those discussed in \citet{Pfahl08} based on the equilibrium tide 
approximation can be present. As more {\it{Kepler}} data become 
available it will be interesting to try to distinguish  between 
these approximations, as they can lead to measuring internal 
properties of the star. For example, in the frame of the  
rotating binary, the star is spinning retrograde if 
$P_{spin} < P_{orb}$, which is expected to be true for most hot 
Jupiter planets. This could induce a phase shift in the 
ellipsoidal variations, as the star's spin ``drags'' the tide 
away from the line of centers. If a phase-lag not associated 
with the planet light can be unambiguously measured, it may be 
possible to put constraints on the tidal Q factor. This 
asymmetry could also lead to interesting orbital dynamics 
effects. 
The detailed shape and amplitude of the ellipsoidal variation 
also depends on the stellar envelope (convective vs. radiative -- 
\citet{Pfahl08}), thus potentially offering a probe of stellar 
interiors. However, for HAT-P-7, the situation is complicated by 
the fact that the spin axis is not aligned with the orbit axis 
\citep{Winn09}, so these optimistic statements must be tempered 
with caution.

Fig.~2 shows model light curves that illustrate the contributions 
of the star to the optical light. The lower curve is the stellar 
light only, showing the ellipsoidal variations. It is not 
constant outside of transit, and its amplitude can be significant 
compared to the depth of the occultation. Above this are four 
curves that include the planet contribution. The lowest is for an 
isothermal isotropic planet with no phase-dependent scattered or 
re-radiated emission. To first order, it is essentially an 
additive offset to the star's light, the amount depending on the 
relative radius and temperature of the planet. If the absorbed 
incident energy from the star is not perfectly uniformly 
re-distributed around the planet, the day side of the planet will 
be hotter than the night side, resulting in an orbit-phase 
dependent modulation peaking at the sub-stellar point (phase 
0.5). Or, if the atmosphere scatters the incident radiation, 
again the day side will be brighter than the night side (though 
not necessarily hotter). The upper three curves show this 
phase-dependent effect for different values of the heat albedo, 
which is a proxy for scattered and re-radiated emission on the 
day side. Note that advection can move the peak downwind and 
produce a phase shift in the planet's emission, as seen in the 
infrared phase curve of HD189733 \citep{Knutson07}; this is not 
included in these models.

Given a stellar temperature estimate, the ELC model can yield 
day and night hemisphere temperatures. Because the planet is 
not black it contributes light at all phases (other than 
occultation) and its flux contribution at each orbital phase, 
relative to ellipsoidal and other effects, allows its 
temperature to be estimated. 
In particular, the ability to measure relative temperatures 
arises from the requirement of getting the occultation and 
transit depths correct given the tight geometric constraints 
(e.g. a 2600~K planet reduces the 6000~\AA \ transit depth by 
29~ppm compared to a zero temperature planet). We find the 
peak temperature on the planet (at the sub-stellar point) is 
3160~K, but a more meaningful flux-weighted day temperature is 
2885 $\pm$100~K. This is notably higher than the 2560 $\pm$100~K 
estimate by \citet{Borucki09}. 
We caution that the day-side temperature estimate is derived 
from the assumption that the light is entirely thermal in origin 
with no scattered component; thus this is a maximum day-side 
temperature estimate for the given heat albedo.
We find the night side temperature $T_{p}$ is a surprisingly hot 
2570 $\pm$95~K, only 590~K less than the peak day temperature.

As seen in Fig.~3, the depth of the occultation is 85.8~ppm, and 
the night-side contribution of the planet is 22.1~ppm, giving a 
day-side peak flux enhancement of 63.7~ppm. Combining these flux 
ratios, temperatures, and the relative radii, we can 
attempt to estimate the geometric and Bond albedos --- see 
\citet{Rowe08} for a discussion. The planet's albedo is key to 
the energy balance in the planetary atmosphere and its 
importance for the dayside emission is discussed in several
papers (e.g. \citet{Seager05}, \citet{Burrows06}, \citet{Lopez07}).
The geometric albedo is given by the ratio of planet to star flux
at phase 0.5, scaled by the ratio of surface areas:
$A_{g} = (F_{p}/F_{*}) / (R_{p}/a)^{2}$.
Using the peak planet to star flux of 63.7~ppm yields 
$A_{g}$=0.18.
This agrees with the $<$0.20 value found for the pM-class 
planet CoRoT-1b (\citet{Snellen09}, \citet{Alonso09b}), but is 
significantly higher than some other planets, notably 
$A_{g}<$0.08 for HD~209458b \citep{Rowe08}, and 0.06 
$\pm$0.06 for CoRoT-2b \citep{Alonso09a}.
The equilibrium temperature and Bond albedo are related by the
definition: $T^{4}_{eq} = T^{4}_{*} (R_{*}/2a)^{2} [f(1-A_{B})]$
where $f$ is the redistribution factor and equals 1 for complete 
redistribution or 2 if the incident flux is re-radiated only on 
the day hemisphere. Setting $f$=1 and equating the night-side 
temperature with the equilibrium temperature should in principle 
allow one to solve for the Bond albedo since all the other terms 
are known. However, this fails because the maximum equilibrium 
temperature possible (when $A_{B}$=0) is 2213~K, considerably 
{\it less} than the ELC estimate. 
It may be that our 6000~\AA \ brightness temperature estimate
exceeds the equilibrium temperature simply because the planet 
is not a blackbody and at other wavelengths lower temperatures 
may be measured. 
We speculate on two other possibilities, both related to the 
presence of a third body in the HAT-P-7 system. 
First, the planet may genuinely be hotter then the equilibrium 
temperature due to non-radiative heating, perhaps tidal heating 
due to an encounter with another object 
(recall the $>$86 degree offset between the orbit and stellar 
spin axes --- \citet{Winn09}, \citet{Narita09}).
A more mundane, but perhaps more likely, explanation is that 
light from a third body is contaminating the flux ratios. 
The acceleration term in the radial velocities suggests the 
presence of another body \citep{Winn09}. When more {\it{Kepler}} 
transits are available, we can check for transit timing 
variations and address the issue of potential third light 
contamination.

In closing, the {\it{Kepler}} light of HAT-P-7 curve reveals 
ellipsoidal variations with an amplitude of approximately 37~ppm. 
This is the first detection of ellipsoidal variations in an 
exoplanet host star, and shows the precision {\it{Kepler}} is 
capable of producing at this early stage. For comparison, a 
transit of an Earth-analog planet around a Sun-like star would 
produce a signal depth of 84~ppm, a factor of 2 larger than this 
effect.

% ..................................................................
\acknowledgments
We thank the anonymous referee for highly valuable comments.
We thank Ron Gilliland for kindly providing assistance with 
the Q1 time series. WFW gratefully acknowledges support from 
Research Corporation for Science Advancement.
The authors acknowledge support from the {\it{Kepler}} 
Participating Scientists Program via NASA grant NNX08AR14G.
Kepler was selected as the 10th mission of the Discovery Program.
Funding for this mission is provided by NASA, Science Mission
Directorate.

{\it Facilities:} \facility{The {\it{Kepler}} Mission}

% \appendix
% ..................................................................

% ..................................................................

\clearpage

\begin{figure}
\includegraphics[scale=0.80,angle=-90]{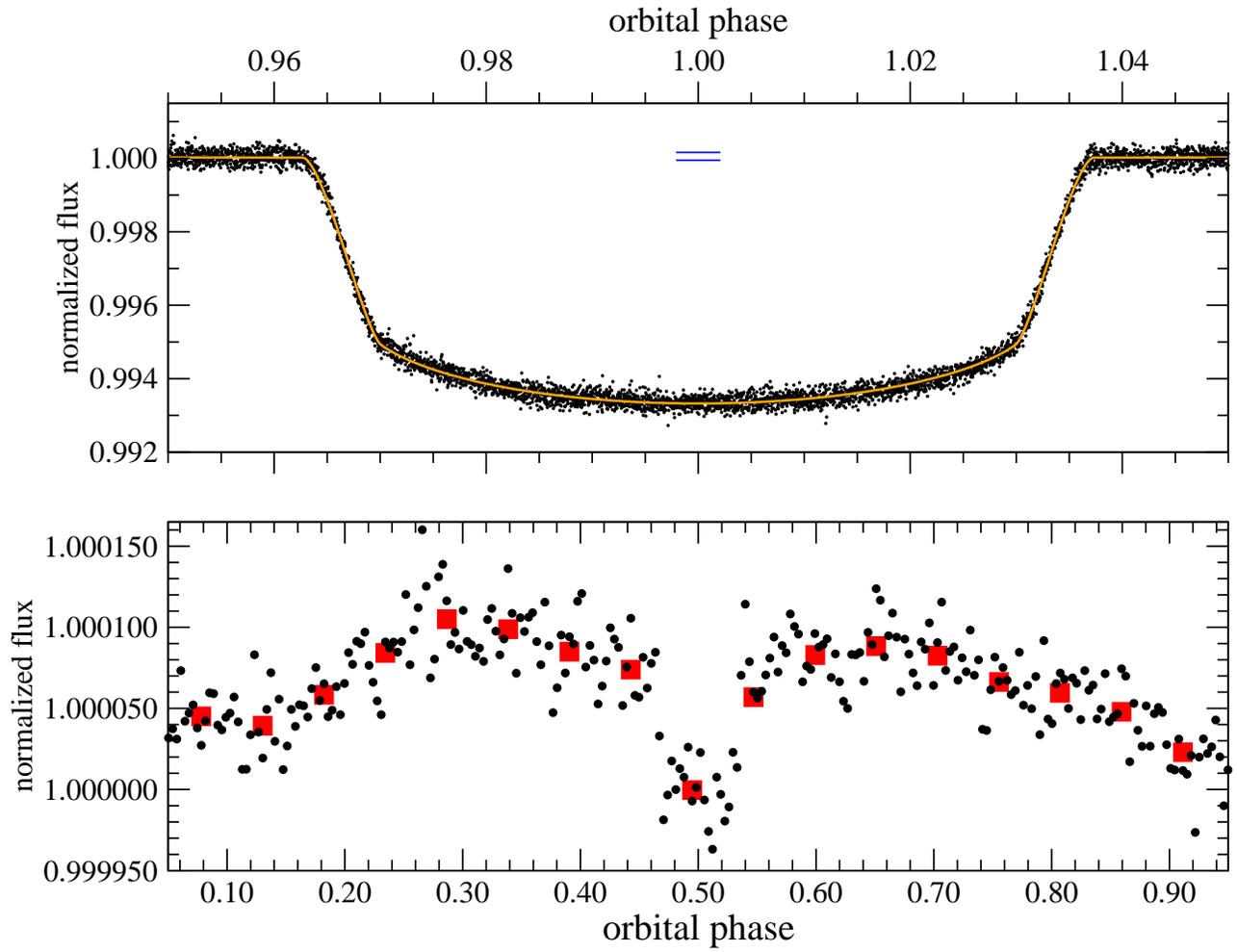}
%\epsscale{0.8}
%\epsscale{1.90}
%\plotone{fig1.ps}
%\plotone{fig1.eps}
\caption{
{\it{upper:}} Detrended and phase-folded light curve at 1-min 
cadence along with the ELC fit. For scale, the horizontal bars 
show the {\it{vertical}} size of the lower panel.
{\it{lower}:}
Light curve averaged in 5~min and 75~min bins.
The double-humped shape is due to the ellipsoidal variations of 
the star plus light from the planet.
\label{fig1}}
\end{figure}

% ................
\clearpage
\begin{figure}
\includegraphics[scale=0.70,angle=-90]{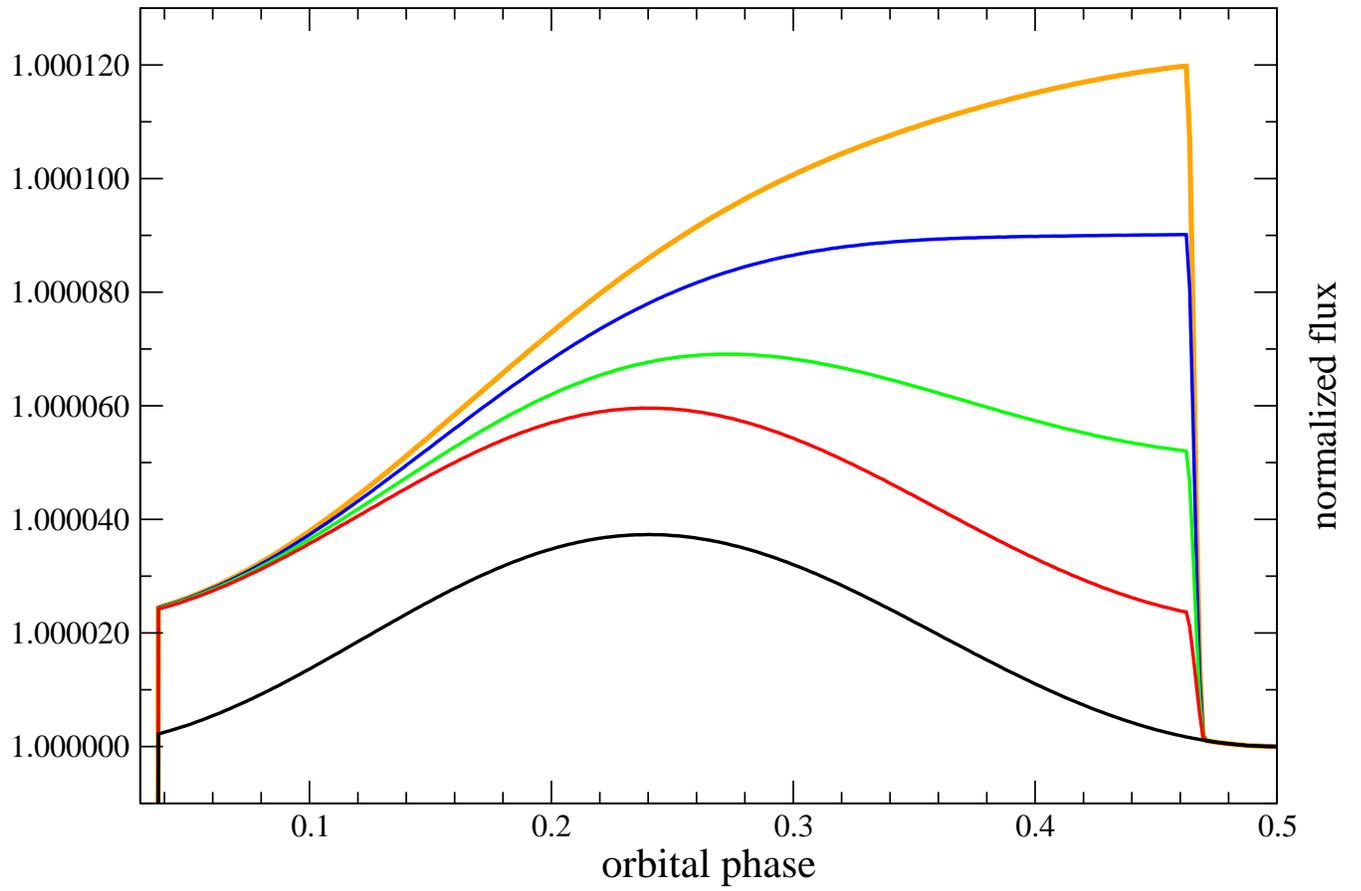}
%\epsscale{0.80}
%\epsscale{1.90}
%\plotone{fig2.ps}
%\plotone{fig2.eps}
\caption{
ELC model light curves for half an orbit. The bottom curve
is the stellar light curve exhibiting the ellipsoidal 
variations. Above this are planet+star light curves
with increasing bolometric (heat) albedos: 0.0, 0.3, 0.6 and 0.8.
If the albedo is 0.0 the planet emits a constant amount of light
resulting in a simple offset in flux. As the albedo increases,
the light from the planet becomes more pronounced.
\label{fig2}}
\end{figure}
% ................

% ................
\clearpage

\begin{figure}
\includegraphics[scale=0.70, angle=-90]{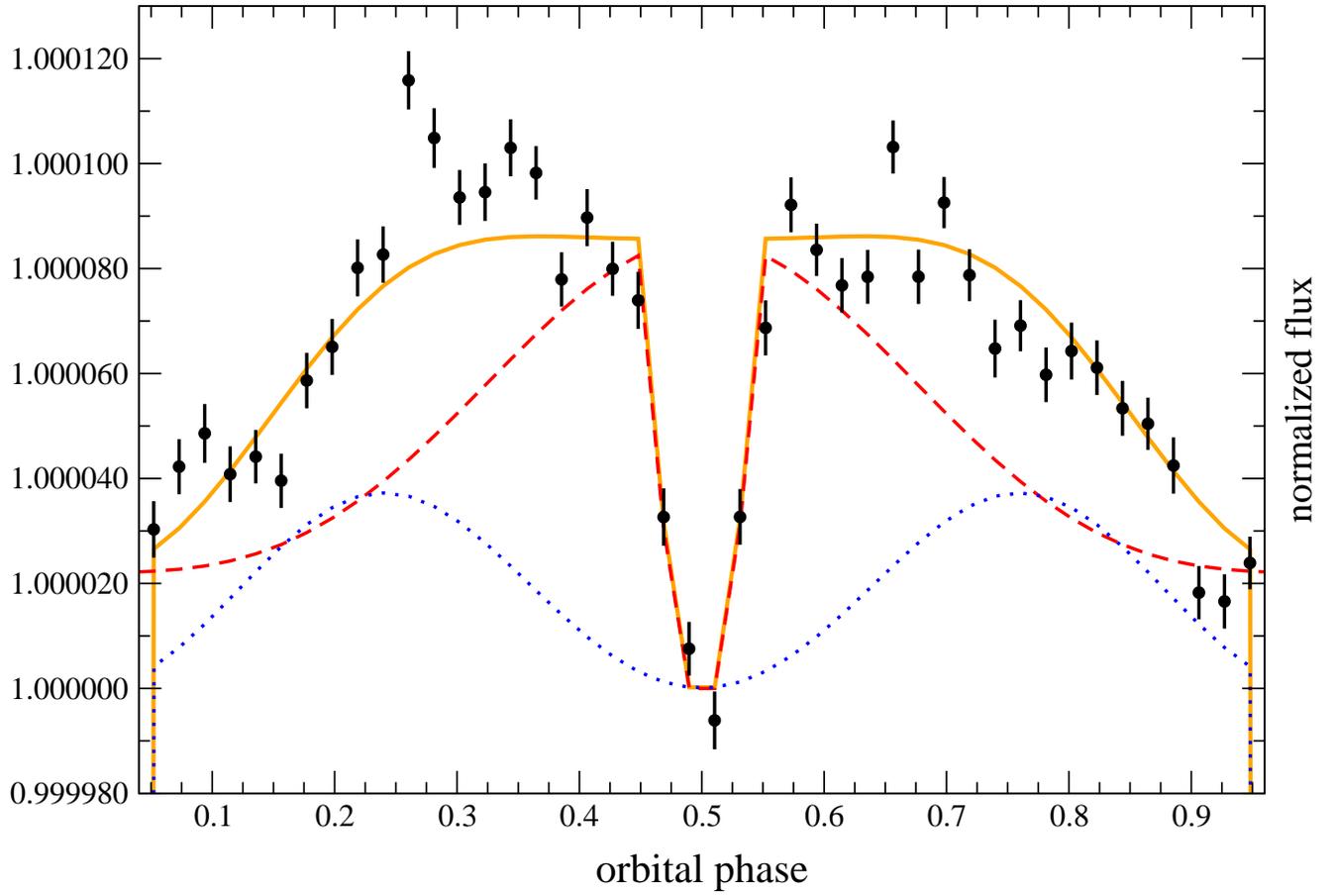}
%\epsscale{0.4}
%\epsscale{1.90}
%\plotone{fig3.ps}
%\plotone{fig3.eps}
\caption{
Phase-folded {\it{Kepler}} light curve with best fit model (solid 
curve). Also shown are the component stellar-only ellipsoidal 
model (dotted) and the planet-only model (offset by +1; dashed). 
Both the data and models have been cast into 30-min bins.
\label{fig3}}
\end{figure}

% ................
\clearpage
\begin{deluxetable}{lrlc}
\tabletypesize{\scriptsize}
\tablecolumns{4}
% \rotate
\tablecaption{HAT-P-7 System Parameters \label{table1}}
\tablewidth{0pt}
\tablehead{
\colhead{Parameter}   & 
\colhead{Value}       & 
\colhead{Uncertainty} & 
\colhead{Unit} 
}
\startdata
T$_{*}$\tablenotemark{a}& 6350   & ---    & K \\
M$_{*}$\tablenotemark{b}& 1.53   & 0.04   & $\Msun$ \\
R$_{*}$\tablenotemark{b}& 1.98   & 0.02   & $\Rsun$ \\
K$_{*}$\tablenotemark{c}& 212    & 5      & $\ms$ \\
\hline
Orbital inclination, i    & 83.1     & 0.5      & degrees \\
Orbital period, P         & 2.204733 & 0.000010 & days \\
Star-to-planet radius, 
              R$_{*}$/Rp  & 12.85    & 0.05   &   \\
limb dark coefficient x   & 0.58     & 0.08   &  \\
limb dark coefficient y   & 0.21     & 0.13   &  \\
Mass of planet, Mp        & 1.82     & 0.03   & $\Mjup$ \\
Radius of planet, Rp      & 1.50     & 0.02   & $\Rjup$ \\
Semimajor axis, a         & 8.22     & 0.02   & $\Rsun$ \\
Bolometric (heat) albedo, 
                A$_{bol}$ & 0.57     & 0.05    &  \\
Tp (night side)           & 2570     & 95      &  K \\
Tp (average day side)     & 2885     & 100     &  K \\
%
% Scaled stellar radius, 
%                a/$R_{*}$  & 4.149  &        &  \\
% Mass ratio, Q=Ms/Mp       & 883    &        &  \\
% Mean density of planet,
%                $\rho_{p}$&         &        & g/cm$^{3}$ \\
% Geometric albedo, A$_{g}$ & 0.18   &        &  \\
%
\enddata
\tablecomments{}
\tablenotetext{a}{Held fixed at 6350 K; \citet{Pal08}}
\tablenotetext{b}{Steered toward  1.52 $\pm$0.036; 
  \citet{Christensen10}}
\tablenotetext{b}{Steered toward 1.991 $\pm$0.018; 
  \citet{Christensen10}}
\tablenotetext{c}{Steered toward 211.8 $\pm$2.6; 
  \citet{Winn09}}
\end{deluxetable}

% ..................................................................
\end{document}